# SCRUB-PA: A Multi-Level Multi-Dimensional Anonymization Tool for Process Accounting

Katherine Luo[‡]    Yifan Li[†‡]    Charis Ermopoulos[†‡]    William Yurcik[‡]    Adam Slagell[‡]

[†]Department of Computer Science
[‡]National Center for Supercomputing Applications (NCSA)
University of Illinois at Urbana-Champaign (UIUC)
{xluo1,yifan,cermopo2,byurcik,slagell}@ncsa.uiuc.edu

## Abstract

In the UNIX/Linux environment the kernel can log every command process created by every user using **process accounting**. This data has many potential uses, including the investigation of security incidents. However, process accounting data is also sensitive since it contains private user information. Consequently, security system administrators have been hindered from sharing these logs. Given that many interesting security applications could use process accounting data, it would be useful to have a tool that could protect private user information in the logs. For this reason we introduce SCRUB-PA, a tool that uses multi-level multi-dimensional anonymization on process accounting log files in order to provide different levels of privacy protection. It is our goal that SCRUB-PA will promote the sharing of process accounting logs while preserving privacy.

## 1.0    Introduction

The UNIX/Linux accounting system collects information on individual/group usage of computer system resources. A system can record every process created by every user. This logged data is called **process accounting** and has been found to be useful for several security purposes [1,6,8], some of which we list below:

- To hold a use accountable for some action indicated in the logs
- To enable the extraction of patterns of use of objects, users or security mechanisms in the system
- To identify security policy violations.
- To create an audit trail of the use (or abuse) that may occur from a specific user.
- To prevent the users from abusing the system by acting as a deterrent, given that the users know that there is a mechanism that logs security relevant actions in the system

While this work is motivated by the sharing of process accounting for security operations, there are other purposes for sharing process accounting data. For example: (1) process accounting has long been the primary data source for usage billing on allocated computer systems, and (2) process accounting data is an accurate source for workload characterization needed to tune applications and schedulers [1,4,5]. However, there are concerns about information disclosure when sharing process accounting data from a specific source since sensitive information will be included in the logs.

We present a tool, SCRUB-PA, to anonymize process accounting log data. Using this tool, the user can anonymize whatever field he/she considers to be sensitive at multiple desired level (e.g., removing all the information, adding noise, or permuting the data). These multi-level anonymization techniques can be employed on different fields—with varying effects on the statistical properties of the overall log.

The rest of the paper is organized as follows. Section 2 presents general background on process accounting. Section 3 provides a description of the general anonymization methods used in SCRUB-PA. In Section 4, we step through all the specific anonymization options in SCRUB-PA for each field in a process accounting record. We conclude with a summary in Section 5.

## 2.0 Background on Process Accounting

### 2.1 Process Accounting

UNIX/LINUX can log every process executed by every user. This kind of logging is called **process accounting** and is generally used in high performance computing (HPC) environments to bill individual users (or groups of users) for the amount of CPU time that they consume [2,7].

Process accounting can be used for security purposes, for instance after a break-in to help determine what commands a user executed, correlating evidence, and incident investigation. Furthermore, process accounting can also be used for other innocuous purposes, such as seeing if anyone is using old unsupported or vulnerable software that should be deleted or determining who is playing games on the fileserver.

Process accounting is performed by the UNIX kernel. Every time a process terminates, the kernel writes a 32-byte record to the **/var/adm/acct** or **/var/adm/pacct** file that includes:

- name of the user and group who created the process
- first eight characters of the name of the command which launched the process
- elapsed time and processor time used by the process
- time that the process exited
- memory usage
- number of disk blocks read or written on behalf of the process
- flags, including:
    - S: Process was executed by the superuser.
    - F: Process ran after a fork, but without an exec.
    - D: Process generated a core file when it exited.
    - X: Process was terminated by signal

Here is the common structure for the process accounting file (in C code):

```
struct acct
  {
    char ac_flag;                       /* Accounting flags.  */
    u_int16_t ac_uid;                   /* Accounting user ID.  */
    u_int16_t ac_gid;                   /* Accounting group ID.  */
    u_int16_t ac_tty;                   /* Controlling tty.  */
    u_int32_t ac_btime;                 /* Beginning time.  */
    comp_t ac_utime;                    /* Accounting user time.  */
    comp_t ac_stime;                    /* Accounting system time.  */
    comp_t ac_etime;                    /* Accounting elapsed time.  */
    comp_t ac_mem;                      /* Accounting average memory usage.  */
    comp_t ac_io;                       /* Accounting chars transferred.  */
    comp_t ac_rw;                       /* Accounting blocks read or written.  */
    comp_t ac_minflt;                   /* Accounting minor pagefaults.  */
    comp_t ac_majflt;                   /* Accounting major pagefaults.  */
    comp_t ac_swaps;                    /* Accounting number of swaps.  */
    u_int32_t ac_exitcode;              /* Accounting process exitcode.  */
    char ac_comm[ACCT_COMM+1];          /* Accounting command name.  */
    char ac_pad[10];                    /* Accounting padding bytes.  */
  };
```

The accounting file **/var/adm/pacct** is accessed by many of the accounting utilities used with system accounting. For example the **lastcomm** or **acctcom** program displays the contents of this file in a human-readable format, as it can be seen in Figure 1.

```
sendmail    F   root            0.05 secs Sat Mar 11 13:28
mail        S   daemon          0.34 secs Sat Mar 11 13:28
send        -   dfr             0.05 secs Sat Mar 11 13:28
post        -   dfr      ttysf  0.11 secs Sat Mar 11 13:28
sendmail    F   root            0.09 secs Sat Mar 11 13:28
sendmail    F   root            0.23 secs Sat Mar 11 13:28
sendmail    F   root            0.02 secs Sat Mar 11 13:28
anno        -   dfr      ttys1  0.14 secs Sat Mar 11 13:28
sendmail    F   root            0.03 secs Sat Mar 11 13:28
mail        S   daemon          0.30 secs Sat Mar 11 13:28
```

**Figure 1.** ASCII Process Accounting Output from *lastcomm* or *acctcom* Utilities

Table 1 shows a list of utilities that display, report, and summarize process information.

**Table 1.** Process Accounting Utilities

| Utility | Description |
| --- | --- |
| ckpacct | Controls the size of the /var/adm/pacct file. |
| monacct | Uses the daily reports created by the utilities above to produce monthly summary reports. |
| shutacct | Records the time accounting was turned off by calling the acctwtmp utility to write a line to the /var/adm/wtmp file. It then calls the turnacct off utility to turn off process accounting. |
| acctcom | Displays process accounting summaries. (available to all users). |
| accton | Turns process accounting on and off. |
| time | Prints real time, user time, and system time required to execute a utility. |
| lastcomm | Displays information about the last processes that were executed. |

The **acctcom** utility is one of the most useful tools for getting a quick report from the system. It can be used to show all the processes that have been executed by a specific user, or to show all the processes, for any user, running longer than **x** seconds etc.

$ acctcom -u (username)
$ acctcom -O 20

System accounting processes generate a number of automated reports that can assist the systems administrator in auditing the daily usage of a system. The Daily Command Summary (DCS) report and Total Command Summary (TCS) report can be found in the **/var/adm/acct/nite** directory and hold several aggregative statistics. The TCS report is stored in an ASCII file called **daycms** and can be viewed with any available text viewer or editor, as shown in Figure 2.

```
TOTAL COMMAND SUMMARY
COMMAND    NUMBER    TOTAL     TOTAL     TOTAL     MEAN      MEAN      HOG       CHARS      BLOCKS
 NAME      CMDS     KCOREMIN  CPU-MIN   REAL-MIN   SIZE-K    CPU-MIN   FACTOR    TRNSFD     READ

TOTALS       82      12.68     0.06      21.91    209.92     0.00      0.28 6.636e+06     0.00
man           1       7.56     0.02       1.68    440.00     0.02      1.02 5.566e+06     0.00
vi            1       2.24     0.02       0.53    121.00     0.02      3.49   71936.00    0.00
ls            5       1.15     0.01       0.02    108.15     0.00     68.33  117144.00    0.00
fgrep        14       0.39     0.00       0.01    124.17     0.00     42.86  286776.00    0.00
tail         14       0.36     0.00       0.02    126.82     0.00     18.03  142744.00    0.00
bsh           6       0.28     0.00       0.01     99.27     0.00     27.50   49410.00    0.00
ps            1       0.21     0.00       0.00    137.00     0.00     66.67   19696.00    0.00
ftpd          1       0.20     0.00       0.81    155.00     0.00      0.16   41576.00    0.00
sendmail      1       0.12     0.00       0.00    468.00     0.00    100.00   13744.00    0.00
fwtmp         3       0.07     0.00       0.00    143.00     0.00     25.00   35840.00    0.00
more          2       0.05     0.00       2.41    195.00     0.00      0.01   30144.00    0.00
pg            3       0.03     0.00      14.78     28.50     0.00      0.01   61232.00    0.00
ksh           2       0.00     0.00       0.00      0.00     0.00      0.00   18360.00    0.00
```

**Figure 2.** Example Summary Process Accounting Report

## 2.2  Process Accounting Limitations

Process accounting data is subject to some inherent limitations with respect to security monitoring. For our purposes, we note two such limitations here. The first limitation is that process accounting does not keep track of parameters passed with the executed command. In fact, it only keeps track of the first eight characters of the command executed. Based on this, a malicious user could link a malicious tool to one with an innocent name and then execute the linked file.

```
$ ln autohack inoccent_prog
$ ./inoccent_prog
```

Furthermore, the file **/var/adm/acct** where all the process accounting data is stored could be altered or even deleted by a malicious attacker who has superuser access, thus removing all evidence of an "attack". This limitation can be addressed by setting up a script running by **cron** that copies the specific file to a different server that would be used as a log data warehouse.

## 3.0  Anonymization Methods

Several anonymization algorithms used in the tool SCRUB-PA are reused multiple times in the context of different process accounting fields. The goal of this section is to introduce detailed descriptions of these general anonymization algorithms here for referral when later in the paper an anonymization method is mentioned in context for a specific process accounting field.

### 3.1  Black Marker Anonymization

In the black marker anonymization method, all the information of the field is deleted. This is done simply by replacing every value of that field with a predefined constant matching the value type expected in the field to be changed. Since all the original information in the field is eliminated, this method results in 100 percent information loss.

### 3.2  Pure Randomization

We use **pure randomization** to describe truly random permutations. This means it is possible create **any** valid permutation. To accomplish this, we make use of tables to store the mappings between unanonymized and the corresponding anonymized values. Because the creation of the table is dependent upon both the state of the PRNG and actual log being anonymized—specifically field values and the order in which they appear—mappings will be different every time this algorithm is run. If mappings must be consistent between different logs that are anonymized, we recommend the use of **keyed randomization**.

## 3.3 Keyed Randomization

With **keyed randomization** the mapping from unanonymized to anonymized data is well-defined by a small key. Thus mappings are consistent between different logs anonymized with this method at different times—as long as the same key is used. We implement this method with keyed hashes (Also, called HMACs). The result of this implementation is that the data is no longer permuted. However, collisions are low, and hence it is nearly a one-to-one correspondence between unanonymized and anonymized values.

## 3.4 Grouping

In the grouped anonymization method, every original value is replaced with a label value representing one predefined subset group to which the original value belongs. In other words, the complete set of values is partitioned, and a canonical member of each equivalence class is chosen to replace all values from that equivalence class. The equivalence classes in the partition do not need to be of equal cardinality and can be formed arbitrarily by developers as long as the user is informed of the partitioning method within SCRUB-PA. This method makes four assumptions (1-3 are just the definition of a partition):

1. the range of all possible original values is known,
2. the union of all subset groups contains all possible values,
3. the subsets chosen are mutually exclusive, and
4. the representative label value for each subset must be an element of that subset.

For example, if the range of all possible original values are the integers 1–30 inclusive, then we could partition the set into the following subsets: subset $A = [1–10]$, subset $B = [11–20]$, and subset $C = [21–30]$. The representative label values for $A$, $B$ and $C$ could be 5, 15, and 25, respectively.

## 3.5 Truncation

In the truncation method, each original value is replaced with the value truncated at a predefined truncation point. To truncate a value, we simply delete everything after the truncation point. Note, truncation is really just a special type of grouping where subgroups are created by truncating. Furthermore, all of these subsets are the same size for a given truncation point. Truncation is a flexible method since it allows the user to choose the amount of information that she wants to reveal by selecting the most appropriate truncation point. The more truncation points available to the user, the finer the granularity of control given to the user.

For example, assume the original data are timestamps in the form of **year:month:day:hour:minute:second**. Truncation points are logical in between the year and month, the month and day, the day and hour, the hour and minute, and the minute and second. If the original value is: **2005:12:21:17:55:07**

truncation at seconds yields: **2005:12:21:17:55:00**
truncation at minutes yields: **2005:12:21:17:00:00**
truncation at hours yields: **2005:12:21:00:00:00**
truncation at days yields: **2005:12:00:00:00:00**
truncation at months yields: **2005:00:00:00:00:00**
truncation at years yields: **0000:00:00:00:00:00**

## 3.6 Random Shift

With the random shift method, every original value is replaced with a new value that is the original value plus a random number from a predefined range. We assume that is the data can be assigned numerical values, and shift numbers are not allowed that overflow the field. The user must select the upper and the lower limits of the random shift. One achieves more security using a larger window and thus choosing from a larger pool of numbers by which to shift. However, large shift numbers may make data look unrealistic. For example, a random shift in time may put timestamps a million years into the future. If one does not want to reveal anonymization has been performed, they must create an upper limit. Note, since we control the shift number as not to overflow a field, order is always preserved.

### 3.7 Enumeration

The enumeration method applies exclusively to timestamps. The result is that all temporal information is removed except for the order in which the events occurred. This means that unsorted logs must be sorted chronologically. Since perfect sorting on a large log is slow and sorting a streamed log is impossible, we make an approximation that works as long as logs are never terribly disordered. This is a reasonable assumption as most logs are sorted and disorder occurs only because of clock skew between sensors and travel time of alerts over the network. We implement this local sorting through the use of a **sliding window** based sorting algorithm.

The size of this window determines the amount of entries to locally buffer and sort. After each local sort within the window is performed, the window advances one place in the log and another local sort is performed. This continues until the window hits the end of the log. Suppose that a log is known to have no out of order entries further than **n** records apart. In this case, if record A occurs more than **n** entries before record B, then we know A is less than B. Such a log could be sorted perfectly using a window of size **n**. In general, one does not know the maximum distance between two out of order entries, but an upper bound can be assumed or estimated. Even if one cannot estimate, or the bound is too large, they can chose the buffer size to trade-off between accuracy and algorithm speed.

### 3.8 Default Settings and Selecting/Unselecting Anonymization Methods

When a field is selected to be anonymized, SCRUB-PA sets the top anonymization option as the default option. The anonymization option for each field is confirmed by the user clicking the OK button at the bottom of the window. If no anonymization (no change) is decided upon even though the anonymization window is open, the user can select no change by clicking on the X in the upper righthand corner of the window to close the window. If an anonymization option is selected for a particular field and later the user changes his/her mind then this selection can be reversed by clicking again on the selected field to un-select anonymization of this field.

## 4.0 Using SCRUB-PA

SCRUB-PA can be downloaded from
<http://security.ncsa.uiuc.edu/distribution/Scrub-PADownLoad.html>
where there is documentation on installation.

### 4.1 Input / Output

The first step in using SCRUB-PA is selecting an input file containing process accounting data. Figure 3A shows the main menu with the input file selection window highlighted. The input file name can be typed in or filled in by selecting it through the file browser. The output file can be designated similarly. Figure 3A also shows the main menu with the output file selection window highlighted. If an existing file is selected then SCRUB-PA will verify with a user window whether the file should be overwritten.

Figure 3B shows the task summary window that appears immediately after the anonymization tasks are complete and an output file has been created. The task summary window documents all the anonymization operations and parameters performed on the input file resulting in the output file (except for the seed keys which are not documented). This metadata about anonymization operations on a specific input-to-output file transaction can be saved or printed to provide a record for future reference since no metadata is kept in the processing accounting files themselves.

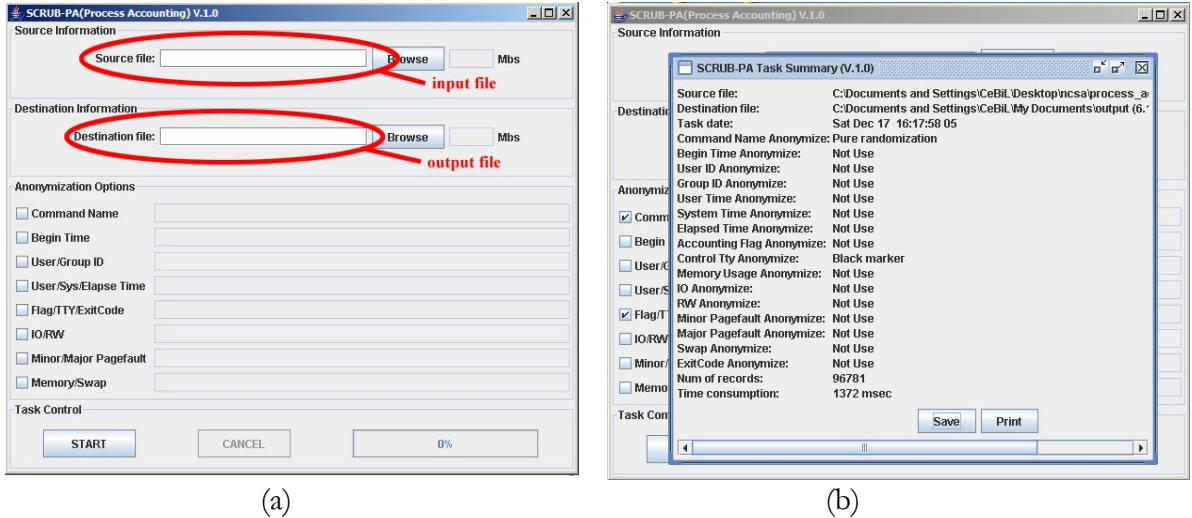

**Figure 3.** SCRUB-PA Main Graphical User Interfaces:
(a) main menu with file I/O fields highlighted
(b) task summary window after anonymization is complete

The remainder of this section will go into depth describing the anonymization options available for each process accounting field.

## 4.2   Command Name Anonymization [Field 1]

The first field in a processing account record for potential anonymization is the "Command Name" field. While this field is potentially the most valuable field for analyzing user behavior (when mapped to User IDs or Group IDs), there may be situations when it is desirable to look at other behavior (such as time usage) independent of what commands are executed. When the Command Names are mapped to User IDs and Group IDs, this field potentially has the most sensitive information since it reveals behavior. This field is defined as follows:

ac_comm: command name

The "Command Name" field has four anonymization options: (1) Black Marker, (2) Pure Randomization, (3) Keyed Randomization, and (4) Grouping. Figure 4 shows the Command Name field's GUI for anonymization options. By clicking in the circle next to the selected anonymization option, the information about that option will become highlighted/darken while the other unselected options will become faded. Anonymization options for each field are mutually exclusive (no more than one anonymization option per field may be selected). Although one anonymization option is selected as a default for each field (usually the first option) no anonymization option needs to be selected.

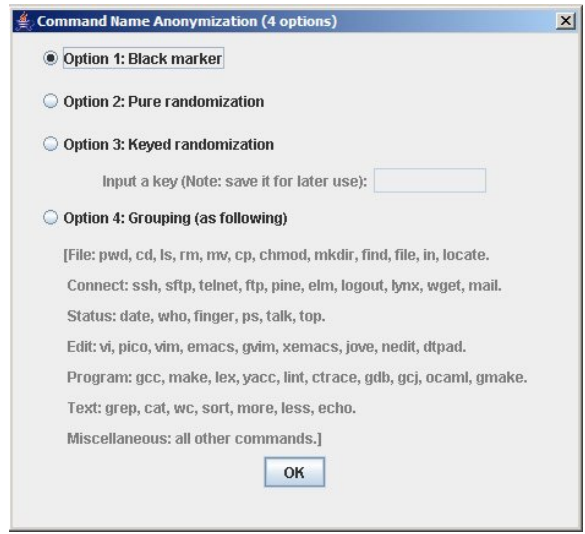

**Figure 4.** Command Name Anonymization GUI

### "Command Name" Field Anonymization Option (1): Black Marker
Using this method, our tool replaces every command name from the log with the label "command".

### "Command Name" Field Anonymization Option (2): Pure Randomization
In this method, our tool replaces every command name with the label "COMMxxxx", where xxxx is a distinct integer. The first command in the log will be replaced by "COMM1", the second by "COMM2", and so on.

### "Command Name" Field Anonymization Option (3): Keyed Randomization
The keyed randomization option will initially ask the user for a key and then map the original command value in the field to a randomly generated string using the provided key as seed. Note that even though it involves randomization, the same input will always get mapped to the same output when using the same key. Essentially, this is a keyed hash.

### "Command Name" Field Anonymization Option (4): Grouping
Having partitioned all linux/unix commands into seven different groups based on their functionality, this anonymization method replaces each command name with the label of the corresponding group that each command belongs to. Below we list the seven groups of commands that we used.

**Table 2.** Group Subset Labels for Unix Commands

| "File" | "Connect" | "Edit" | "Program" | "Text" | "Status" | "Miscellaneous" |
|---|---|---|---|---|---|---|
| pwd | ssh | vi | gcc | grep | date | "all other commands that does not fall in one of the previous categories" |
| cd | sftp | pico | make | cat | who | |
| ls | telnet | vim | lex | wc | finger | |
| rm | ftp | emacs | yacc | sort | ps | |
| mv | pine | gvim | lint | more | talk | |
| cp | elm | xemacs | ctrace | less | top | |
| chmod | logout | jove | gdb | echo | | |
| mkdir | lynx | nedit | gcj | | | |
| rmdir | wget | dtpad | ocaml | | | |
| find | mail | | gmake | | | |
| file | | | | | | |
| in | | | | | | |
| locate | | | | | | |

## 4.3 Begin Time Field Anonymization   [Field 2]

The second field in a process accounting record for potential anonymization is the "Begin Time" Field. This field is defined as follows:

**ac_btime: beginning system time for the process**

The "Begin Time" field is potentially a valuable analysis field for correlating process behavior with the time certain events occurred. However, there may be situations when correlating behavior with times reveals private information such that anonymization at different levels is desirable.

The "Begin Time" field has three anonymization options: (1) Time Unit Annihilation, (2) Random Time Shift, and (3) Enumeration. Figure 5 shows the "Begin Time" field's GUI for anonymization options. By clicking in the circle next to the selected anonymization option, the information about that option will become highlighted/darken while the unselected options will become faded. Anonymization options for each field are mutually exclusive (no more than one anonymization option per field may be selected). Although one anonymization option is selected as a default for each field (usually the first option) no anonymization option needs to be selected.

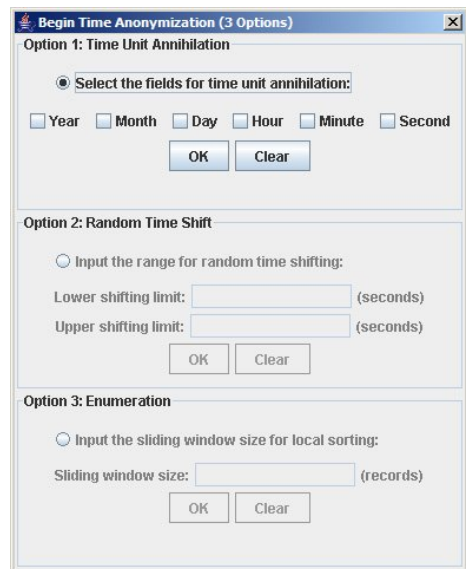

**Figure 5.** Begin Time Anonymization GUI

### "Begin Time" Field Anonymization Option (1): Time Unit Annihilation
Using this anonymization method, any time component (year, month, day, hour, minute, or second) can be annihilated—set to zero. This is a generalization of truncation which would always make the granularity of time rougher. Note, by selecting odd combinations of values (e.g., month and hour), you could make it so records no longer appear to be in order.

### "Begin Time" Field Anonymization Option (2): Random Time Shift
In some cases it may be important to know how far apart two events are temporally without knowing exactly when they happened. For this reason we introduce the random time shift anonymization method, where all timestamps are shifted by a random variable. The user can set the upper and the lower limits of the random variable. This is mostly to prevent times that occur far in the future and would indicate anonymization had been done. For example, a purely random integer would likely send the dates millennia into the future.

### "Begin Time" Field Anonymization Option (3): Enumeration

In this anonymization method the records are sorted into sequential timestamp order using a local sort window set by the user (See details of this sliding window implementation in section 3.6). Inter-arrival time between processes are not preserved, only the sequence. In fact, even the start time of the first record is not preserved.

## 4.4 "User ID"/"Group ID" Anonymization [Fields 3,4]

The third and fourth fields in a process accounting record for potential anonymization are the "User ID" and "Group ID" Fields. These fields are defined as follows:

> ac_uid: user ID associated with the process
> ac_gid: group ID associated with the process

While these ID fields are potentially the most valuable for analyzing user behavior, there may be situations when it is desirable to look at other behavior (such as command usage) independent of what specific users are involved. Thus, it is prudent, for privacy's sake, to anonymize these fields when individuals users do not need to be identified. Note that depending on the anonymization option chosen, one can still correlate all actions by a particular user, they just will be unable to determine who that user is.

Figure 6 shows the "User ID" / "Group ID" fields' GUI for anonymization options. Either ID Field can be anonymized independently by clicking on the box next to each Field. By clicking in the circle next to the selected anonymization option, the information about that option will become highlighted/darken while the unselected options will become faded. Anonymization options for each field are mutually exclusive (no more than one anonymization option per field may be selected). Although one anonymization option is selected as a default for each field (usually the first option) no anonymization option needs to be selected.

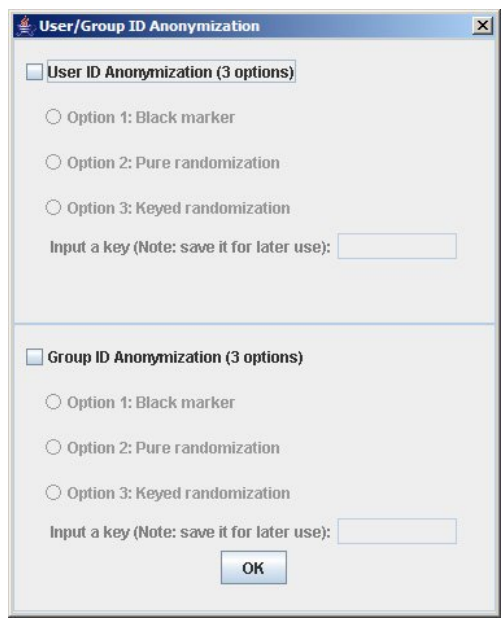

**Figure 6.** User ID/Group ID Anonymization GUI

Each of these ID fields has three anonymization options: (1) Black Marker, (2) Pure Randomization, and (3) Keyed Randomization.

### "User ID"/"Group ID" Fields Anonymization Option (1): Black Marker
The black marker option will replace this field with a zero.

### "User ID"/"Group ID" Fields Anonymization Option (2): Pure Randomization
The pure randomization option will replace the input value with a random variable from 0 to 65535. In other words, a permutation on the set of all possible IDs is performed. Thus data can be correlated to a particular user/group even though the true identity of the user/group is unknown. Since this algorithm is table based, anonymization between different logs done at different times will have different mappings.

### "User ID"/"Group ID" Fields Anonymization Option (3): Keyed Randomization
The keyed randomization option will initially ask the user for a key and then map the original value in the field to a random value [0…65635] using the provided key as seed. This is done through keyed hash algorithm. This means that even though the same user/group will always map to the same value (using a fixed key), the converse is not true. In other words, it is possible that two or more IDs map to the same value since the hash function is not injective. For less than 256 distinct users, such a collision is statistically unlikely, though.

## 4.5 Time Usage (User Time/System Time/Elapsed Time) Anonymization [Fields 5,6,7]

The fifth, sixth, and seventh fields in a process accounting record for potential anonymization are the "User Time", "System Time", and "Elapsed Time" fields, collectively called the Time Usage Fields and defined as follows:

> ac_utime: time spent in user space for the process
> ac_stime: time spent in kernel space for the process (accurate to 0.01 second)
> ac_etime: total elapsed time for the process (greater than or equal to ac_utime + ac_stime)
> also known as "wall clock" or real-time (accurate to a second)
>
> {Note: these fields are reported in seconds with different accuracy}

The Time Usage Fields are the most valuable field for resource accounting and billing when mapped to User IDs or Group IDs, however, there may be situations when it is desirable not to reveal resource consumption. In general, the anonymization of any of these fields may destroy the relationship between them. It is known that User Time + System Time $\geq$ Elapsed Time. Since the relationship between the fields may be destroyed after anonymization, the SCRUB-PA tool informs the user of this possibility through a pop-up window.

Figure 7 shows the Time Usage fields' GUI for anonymization options. Each Time Usage Field can be anonymized independently by clicking on the box next to each Field. Then by clicking in the circle next to the selected anonymization option, the information about that option will become highlighted/darken while the unselected options will become faded. Anonymization options for each field are mutually exclusive (no more than one anonymization option per field may be selected). Although one anonymization option is selected as a default for each field (usually the first option) no anonymization needs to be used.

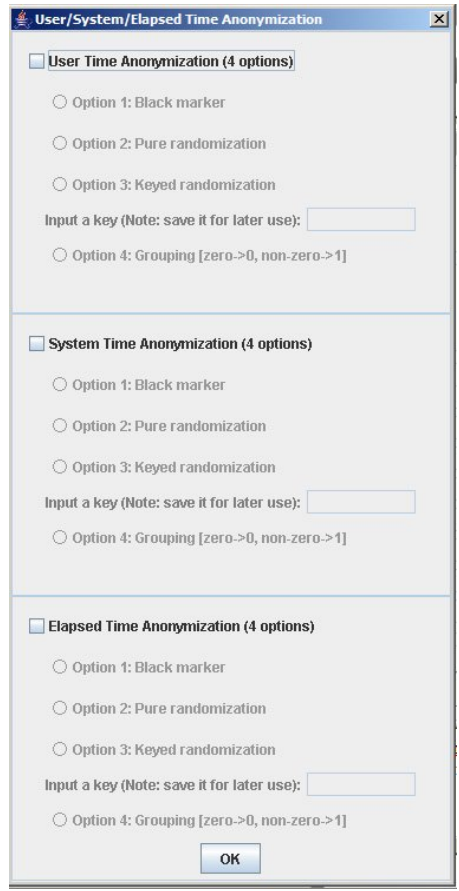

Figure 7. Time Usage Anonymization GUI (User Time/System Time/Elapsed Time)

Each of the Time Usage Fields has four anonymization options: (1) Black Marker, (2) Pure Randomization, (3) Keyed Randomization, and (4) Grouping.

### "Time Usage" Field Anonymization Option (1): Black Marker
The black marker option will replace this field with a zero.

### "Time Usage" Field Anonymization Option (2): Pure Randomization
The pure randomization option will replace the input value with a random value from 0 to 65535. This is done by creating a random permutation. Thus, a time of **x** seconds that maps to **y** seconds will do so for every entry in the log. However, these mappings will likely change between different runs on different logs. Note that relationships such as one process taking longer than another are completely destroyed.

### "Time Usage" Field Anonymization Option (3): Keyed Randomization
The keyed randomization option will initially ask the user for a key and then map the original value in the field to a random value using the provided key as a seed. We implement this through a type of keyed hash. Thus, this mapping remains consistent through runs on different logs as long as the same key is used. Note that relationships such as one process taking longer than another are completely destroyed. Furthermore, since a hash is used on a small space of values, collisions are likely.

### "Time Usage" Field Anonymization Option (4): Grouping
This option maps all zero input values to zero and all non-zero input values to one.

## 4.6 Accounting Flag/Controlling TTY/ExitCode Anonymization [Fields 8,9,10]

The eighth, ninth, and tenth fields in a process accounting record for potential anonymization are the "Accounting Flag", "Controlling TTY", and "ExitCode" fields. These fields are defined as follows:

>ac_flag: accounting flags
>>S: process was executed by the superuser
>>F: process ran after a fork, but without exec
>>D: process generated a core file when it exited
>>X: process was terminated by signal
>
>ac_tty: TTY which sent the command to launch the process
>
>ac_exitcode: a status code that indicates how the process terminated
>>EXIT_SUCCESS
>>EXIT_FAILURE

The "Accounting Flag" and "Controlling TTY" fields are potentially valuable fields for analysis since they may provide information correlating a command to a user or process. However, there may be situations when correlating behavior reveals private information such that anonymization at different levels is desirable. The "ExitCode" field reveals behavior about the process itself which may or may not correlate to private information, however, a mechanism for anonymizing this field is provided for cases when it is desired.

Figure 8 shows the "Accounting Flag" / "Controlling TTY" / "ExitCode" fields' GUI for anonymization options. Each of these fields may be anonymized independently by clicking on the box next to each Field. Then by clicking in the circle next to the selected anonymization option, the information about that option will become highlighted/darken unselected options will become faded. Anonymization options for each field are mutually exclusive (no more than one anonymization option per field may be selected). Although one anonymization option is selected as a default for each field (usually the first option) no anonymization option needs to be used.

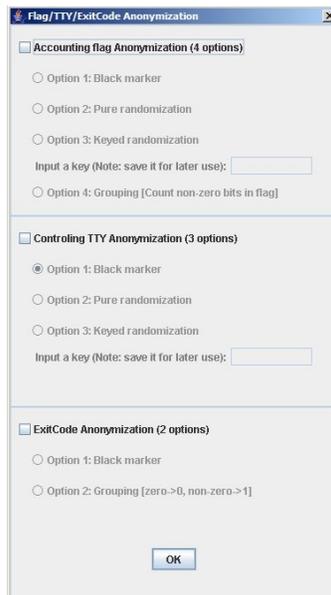

**Figure 8.** "Accounting Flag"/"Controlling TTY"/"Exitcode" Anonymization GUI

The "Accounting Flag" field has four anonymization options: (1) Black Marker, (2) Pure Randomization, (3) Keyed Randomization and (4) Grouping.

### "Accounting Flag" Field Anonymization Option (1): Black Marker
The black marker option will replace this field with a zero.

### "Accounting Flag" Field Anonymization Option (2): Pure Randomization
For the "Accounting Flag" field, input values are replaced with a random number from 0 to 255 masked with 0x0001011. This creates a permutation on the set of possible values that is dependent on the internal state of the PRNG and the details of the log file.

### "Accounting Flag" Field Anonymization Option (3): Keyed Randomization
The keyed randomization option will initially ask the user for a key and then map the original value in the field to a random value using the provided key as seed. This is implemented through a type of keyed hash. Because of the small input space, hash collisions are likely. However, mappings are the same between different logs if the same key is used.

### "Accounting Flag" Field Anonymization Option (4): Grouping
For the "Accounting Flag" field, the input is replaced by an integer representing the number of non-zero bits in the original value.

The "Controlling TTY" field has three anonymization options: (1) Black Marker, (2) Pure Randomization, and (3) Keyed Randomization.

### "Controlling TTY" Field Anonymization Option (1): Black Marker
The black marker option will replace this field with a zero.

### "Controlling TTY" Field Anonymization Option (2): Pure Randomization
For the TTY field, input values are replaced with a random number from 0 to 65535. This is done by creating a random permutation. However, these mappings will likely change between different runs on different logs.

### "Controlling TTY" Field Anonymization Option (3): Keyed Randomization
The keyed randomization option will initially ask the user for a key and then map the original value in the field to a random value using the provided key as seed. This is implemented through a type of keyed hash. Hash collisions are unlikely if there are less than 256 distinct TTYs in a log. Also, mappings are the same between different logs if the same key is used.

The "ExitCode" field has two anonymization options: (1) Black Marker and (2) Grouping.

### "ExitCode" Field Anonymization Option (1): Black Marker
The black marker option will replace this field with a zero.

### "ExitCode" Field Anonymization Option (2): Grouping
For the "ExitCode" field, all non-zero values are mapped to one and zero values are left zero.

## 4.7 Chars Transferred/Blocks Read or Written Anonymization [Fields 11,12]

The eleventh and twelfth fields in a process accounting record for potential anonymization are the "Chars Transferred" and the "Blocks Read or Written" fields. These fields are defined as follows:

> ac_io: number of characters transferred by the process

> ac_rw: number of blocks read or written by the process

The "Chars Transferred" and "Blocks Read or Written" fields reveal behavior about the process itself which may or may not correlate to private information, however, mechanisms for anonymizing these fields are provided for cases when it is desired.

Figure 9 shows "Chars Transferred" / "Blocks Read or Written" fields' GUI for anonymization options. Each of these fields may be anonymized independently by clicking on the box next to each Field. Then by clicking in the circle next to the selected anonymization option, the information about that option will become highlighted/darken while the other options not selected will become faded. Anonymization options for each field are mutually exclusive (no more than one anonymization option per field may be selected). Although one anonymization option is selected as a default for each field (usually the first option) no anonymization needs to be done.

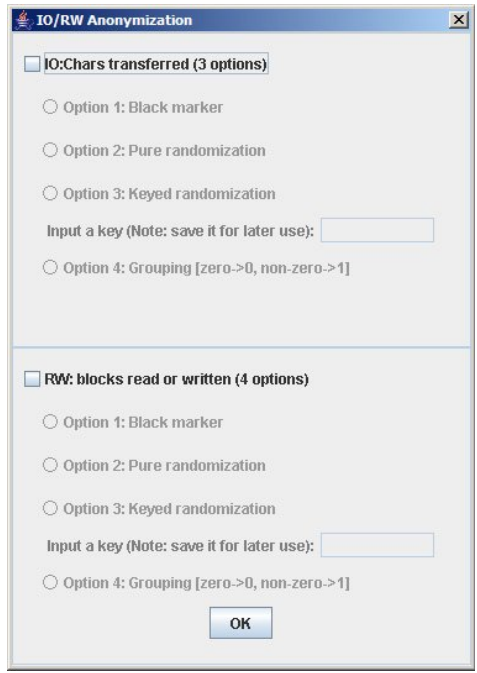

**Figure 9.** "Chars Transferred"/"Blocks Read or Written" Anonymization GUI

The "Chars Transferred" field has four anonymization options: (1) Black Marker, (2) Pure Randomization, (3) Keyed Randomization, and (4) Grouping.

### "Chars Transferred" Field Anonymization Option (1): Black Marker
The black marker option will replace this field with a zero.

### "Chars Transferred" Field Anonymization Option (2): Pure Randomization
The pure randomization option will create a random permutation on the set of possible values. This random permutation is dependent upon the state of the PRNG and the log file itself. Thus, the mapping will likely change for different logs.

### "Chars Transferred" Field Anonymization Option (3): Keyed Randomization
The keyed randomization option will initially ask the user for a key and then map the original value in the field to a random value using the provided key as seed. This is implemented through a keyed hash. This means that logs anonymized at different times use the same mapping if the same key is used.

### "Chars Transferred" Field Anonymization Option (4): Grouping
This options maps every non-zero value to one, while every zero value is left zero.

The "Blocks Read or Written" fields has four anonymization options: (1) Black Marker, (2) Pure Randomization, (3) Keyed Randomization, and (4) Grouping.

### "Blocks Read or Written" Field Anonymization Option (1): Black Marker
The black marker option will replace this field with a zero.

### "Blocks Read or Written" Field Anonymization Option (2): Pure Randomization

The pure randomization option will create a random permutation on the set of possible values. This random permutation is dependent upon the state of the PRNG and the log file itself. Thus, the mapping will likely change for different logs.

### "Blocks Read or Written" Field Anonymization Option (3): Keyed Randomization

The keyed randomization option will initially ask the user for a key and then map the original value in the field to a random value using the provided key as seed. This is implemented through a keyed hash. This means that logs anonymized at different times use the same mapping if the same key is used.

### "Blocks Read or Written" Field Anonymization Option (4): Grouping

This options maps every non-zero value to one, while every zero value is left zero.

## 4.8    Minor Pagefaults/Major Pagefaults Anonymization [Fields 13,14]

The thirteenth and fourteenth fields in a process accounting record for potential anonymization are the "Minor Pagefaults" and the "Major Pagefaults" fields. These fields are defined as follows:

> ac_minflt: number of minor pagefaults. Minor pagefaults do not require physical I/O. Any process doing a read or write to something that is in the page will get a minor page fault.

> ac_majflt: number of major page faults. Major pagefaults cause pages to be read from disk including pages read from buffer cache.

The "Minor Pagefaults" and "Major Pagefaults" fields reveal behavior about the process itself which may or may not correlate to private information. However, mechanisms for anonymizing these fields are provided for cases when it is desired.

Figure 10 shows "Minor Pagefaults"/"Major Pagefaults" fields' GUI for anonymization options. Each of these fields may be anonymized independently by clicking on the box next to each Field. Then by clicking in the circle next to the selected anonymization option, the information about that option will become highlighted/darken while the unselected options will become faded. Anonymization options for each field are mutually exclusive (no more than one anonymization option per field may be selected). Although one anonymization option is selected as a default for each field (usually the first option) no anonymization needs be performed.

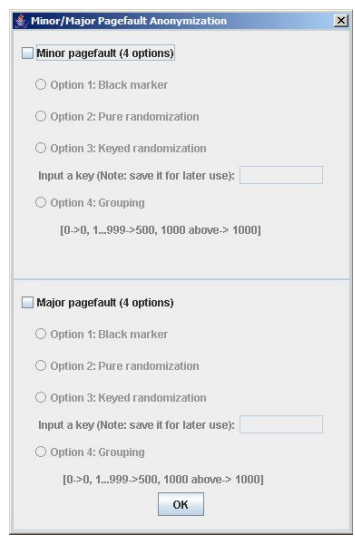

**Figure 10.** "Minor Pagefaults"/"Major Pagefaults" Anonymization GUI

The "Minor Pagefaults"/"Major Pagefaults" fields have four anonymization options: (1) Black Marker, (2) Pure Randomization, (3) Keyed Randomization, and (4) Grouping.

### "Minor/Major Pagefaults" Field Anonymization Option (1): Black Marker
The black marker option will replace this field with a zero.

### "Minor/Major Pagefaults" Field Anonymization Option (2) Pure randomization
The pure randomization option will create a random permutation on the set of possible values. This random permutation is dependent upon the state of the PRNG and the log file itself. Thus, the mapping will likely change for different logs.

### "Minor/Major Pagefaults" Field Anonymization Option (3): Keyed Randomization
The keyed randomization option will initially ask the user for a key and then map the original value in the field to a random value using the provided key as seed. This is implemented through a keyed hash. This means that logs anonymized at different times use the same mapping if the same key is used.

### "Minor/Major Pagefaults" Field Anonymization Option (4): Grouping
Using the predefined groups seen in Table 3, our tool replaces every input value with the corresponding label of the group that it belongs.

**Table 3.** Minor Pagefaults/Major Pagefaults Groups

| Input: Value Range | Output: Group Subset Labels |
|---|---|
| 0 | 0 |
| 1…999 | 500 |
| 1000 and above | 1000 |

## 4.9 Memory Usage/Number of Swaps Anonymization [Fields 15,16]

The fifteenth and sixteenth process fields in a accounting record for potential anonymization are the "Memory Usage" and the "Number of Swaps" fields. These fields are defined as the following:

> **ac_mem:** average amount of memory in units of 8K (pages) that is used by the process.
>
> **ac_swaps:** number of memory swaps

The "Memory Usage" and "Number of Swaps" fields reveal behavior about the process itself which may or may not correlate to private information. However, mechanisms for anonymizing these fields are provided for cases when it is needed.

Figure 11 shows "Memory Usage" / "Number of Swaps" fields' GUI for anonymization options. Each of these fields may be anonymized independently by clicking on the box next to each Field. Then by clicking in the circle next to the selected anonymization option, the information about that option will become highlighted/darken while the unselected options will become faded. Anonymization options for each field are mutually exclusive (no more than one anonymization option per field may be selected). Although one anonymization option is selected as a default for each field (usually the first option) no anonymization needs be done.

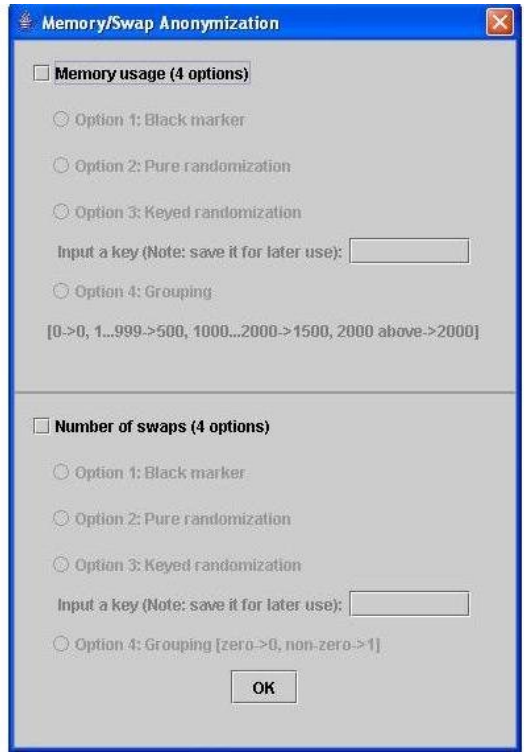

**Figure 11.** Memory Usage/Number of Swaps Anonymization GUI

The "Memory Usage" field has four anonymization options: (1) Black Marker, (2) Pure Randomization, (3) Keyed Randomization, and (4) Grouping.

### "Memory Usage" Field Anonymization Option (1): Black Marker
The black marker option will replace this field with a zero.

### "Memory Usage" Field Anonymization Option (2): Pure Randomization
The pure randomization option will create a random permutation on the set of possible values. This random permutation is dependent upon the state of the PRNG and the log file itself. Thus, the mapping will likely change for different logs.

### "Memory Usage" Field Anonymization Option (3): Keyed Randomization
The keyed randomization option will initially ask the user for a key and then map the original value in the field to a random value using the provided key as seed. This is implemented through a keyed hash. This means that logs anonymized at different times use the same mapping if the same key is used.

### "Memory Usage" Field Anonymization Option (4): Grouping
Using the predefined groups seen in Table 4, our tool replaces every input value with the corresponding label of the group that it belongs.

**Table 4.** Subset Groupings for Memory Usage Field

| Input: Value Range | Output: Group Subset Labels |
|---|---|
| 0 | 0 |
| 1…999 | 500 |
| 1000…2000 | 1500 |
| 2001 and above | 2000 |

The "Number of Swaps" field has four anonymization options: (1) Black Marker, (2) Pure Randomization, (3) Keyed Randomization, and (4) Grouping.

### "Number of Swaps" Field Anonymization Option (1): Black Marker
The black marker option will replace this field with a zero.

### "Number of Swaps" Field Anonymization Option (2): Pure Randomization
The pure randomization option will create a random permutation on the set of possible values. This random permutation is dependent upon the state of the PRNG and the log file itself. Thus, the mapping will likely change for different logs.

### "Number of Swaps" Field Anonymization Option (3): Keyed Randomization
The keyed randomization option will initially ask the user for a key and then map the original value in the field to a random value using the provided key as seed. This is implemented through a keyed hash. This means that logs anonymized at different times use the same mapping if the same key is used.

### "Number of Swaps" Field Anonymization Option (4): Grouping
This options maps every non-zero value to one, while every zero value is left zero.

## 5.0 Summary

In this paper we describe SCRUB-PA, a tool that performs process accounting log anonymization by allowing a user to select multi-level options for anonymizing each field within a process accounting record. This functionality is motivated by the problem of sharing process accounting logs between organizations while simultaneously preserving the privacy of sensitive information within each log. Realizing organizations have different security environments with different levels of privacy requirements, the multiple anonymization options for each field allow flexibility for selecting anonymization schemes that match different environments. It is our hope that SCRUB-PA becomes a valuable tool for security engineers, system administrators, developers, and researchers who want to share process accounting information without revealing sensitive information. A mailing list for SCRUB-PA support is linked off the SCRUB-PA download webpage: **<http://security.ncsa.uiuc.edu/distribution/Scrub-PADownLoad.html>**.

## References

[1] A. Cockcroft, *Processing Accounting Data into Workloads*, Sun BluePrints, 1999.

[2] K. Gilbertson, "Process Accounting," *Linux Journal*, Dec. 12, 2001.

[3] V. Hazelwood, "Unix Accounting Magic," S*ysAdmin Magazine*, March 1999.

[4] *IEEE International Symposium on Workload Characterization – IISWC (formerly known as the Workshop on Workload Characterization –WWC)*, 2005 IISWC website: <http://www.iiswc.org/iiswc2005/home.html>

[5] D.A. Menasce, "Workload Characterization," *IEEE Internet Computing*, Sept/Oct. 2003.

[6] S. Peisert, "Forensics for System Administrators," *Usenix ;Login*, Vol 30 No 4, August 2005.

[7] A. Tam. *Enabling Process Accounting on Linux HOWTO, version 1.1*, 2001. <http://www.tldp.org/HOWTO/ProcessAccounting/>

[8] W. Yurcik and C. Liu, "A First Step Toward Detecting SSH Identity Theft in HPC Cluster Environments, Discriminating Masqueraders Based on Command Behavior," *1st Intl. Workshop on Cluster Security (Cluster-Sec)* in conjunction with *5th IEEE Intl. Symposium on Cluster Computing and the Grid (CCGrid),* 2005.